# Towards Low Cost Coupling Structures for Short-Distance Optical Interconnections

N. Hendrickx[1], J. Van Erps[2], T. Alajoki[3], N. Destouches[4], D. Blanc[4], J. Franc[4], P. Karioja[3], H. Thienpont[2], P. Van Daele[1]

[1]Ghent University, TFCG Microsystems, Dept. of Information Technology (INTEC),
Technologiepark 914A, B-9052 Ghent, Belgium
[2]Vrije Universiteit Brussel, Dept. of Applied Physics and Photonics (TONA-FirW),
Pleinlaan 2, B-1050 Brussels, Belgium
[3]VTT,Kaitoväylä 1, 90570 Oulu, Finland
[3]Laboratoire Hubert Curien, UMR 5516, Université Jean Monnet
18 Rue Benoit Lauras 42000 Saint-Etienne, France
Contact: nina.hendrickx@intec.ugent.be, Phone: +3292645370, Fax: +3292645374

**Abstract**

*The performance of short distance optical interconnections in general relies very strongly on coupling structures, since they will determine the overall efficiency of the system to a large extent. Different configurations can be considered and a variety of manufacturing technologies can be used. We present two different discrete and two different integrated coupling components which can be used to deflect the light beam over 90° and can play a crucial role when integrating optical interconnections in printed circuit boards. The fabrication process of the different coupling structures is discussed and experimental results are shown. The main characteristics of the coupling structures are given. The main advantages and disadvantages of the different components are discussed.*

**Introduction**

The performance of short distance optical interconnections in general relies very strongly on coupling structures, since they will determine the overall efficiency of the system to a large extent. The integration of the optical interconnection to the board-level is done with the use of a polymer optical layer. The optical layer contains multimode optical waveguides, which guide the light in the plane of the optical layer. The availability of VCSELs and photo-detectors at the targeted wavelength of 850nm, which are placed on top of the optical layer, requires the use of coupling structures. These coupling structures deflect the light beam over 90°, enabling light to be coupled from the VCSEL towards the optical waveguide or from the waveguide towards the photo-detector. Different configurations can be considered to obtain the 90° beam deflection. We will discuss two integrated and two discrete coupling structures. The main advantage of the integrated structures is the fact that the alignment between the waveguides and the coupling component is arranged during the fabrication itself whereas the discrete components have to be inserted into the optical layer and require passive or active alignment for placing the component at the right position. The main disadvantage of the integrated structures is the fact that they have to be compatible with the entire fabrication process, which may include elevated pressures and temperatures, whereas the discrete ones can be fabricated in a separate step and inserted into the optical layer in a later phase. In our case, active alignment is used to arrange the alignment between the couplers and the waveguides. In the next sections, the different coupling structures are discussed. The fabrication process is described and experimental realizations are shown. The main characteristics are also given. The different components are finally compared in an objective way, giving the main advantages and disadvantages.

**Laser ablated micro-mirrors**

Laser ablation is a versatile micro-structuring technology that can be used to pattern the main building blocks of the optical interconnection into the polymer optical layer [1]. The ablation set-up available at Ghent University contains three different laser sources: a KrF excimer (248nm), frequency tripled Nd-YAG (355nm) and a $CO_2$ (9.6μm) laser. The excimer laser beam can be tilted with respect to the sample, which eases the fabrication of angled features. During the processing, the sample is placed on a computer-controlled translation stage which has an accuracy of 1μm.

The laser beam is send through an optical projection system and projected onto the sample. The photon energy is absorbed by the polymer material. As soon as the photon density inside the material exceeds a certain threshold, the photon energy can be used for material decomposition. This includes both photo-thermal and photo-chemical processes and results in the ejection of the decomposed material in the form of an ablation plume. This plume contains both evaporated and non-evaporated particles. The solid particles will fall back to the surface and cause debris, which is off course highly undesirable since it increases the surface roughness of the ablated area. Polymer materials typically show a high absorption in the UV-range allowing for an ablation with a low deposition of debris. The optical layer is thus patterned by physically removing material.

The material used for the optical layer, Truemode Backplane$^{TM}$ Polymer, is highly cross-linked acrylate-based polymer which has excellent optical and thermal properties. The optical layer consists of a cladding-core-cladding stack, where the cladding material has a slightly lower refractive index than the core material. The light can in this way be trapped inside the core layer by means of total internal reflection (TIR). Multimode waveguides are used to guide the light in the plane of the optical layer. The waveguide cores are ablated into the core layer with the KrF excimer laser. Material is removed on both sides of the resulting waveguide core. The waveguides have a cross-section of 50µm×50µm and are on a pitch of 125µm. The propagation loss of the ablated waveguides is 0.12dB/cm at 850nm [2]. A cross-section of an array of ablated waveguides is given in Fig. 1.

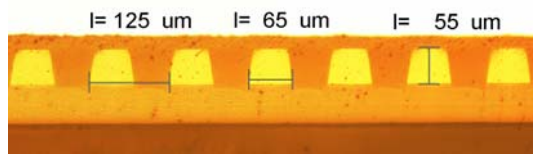

**Figure 1: cross-section of an array of laser ablated multimode waveguides.**

45° micro-mirrors are good candidates for the 90° beam deflection. They are wavelength independent, highly reproducible and can be fabricated with a number of technologies. These 45° micro-mirrors can be ablated into the optical layer with use of the tilted KrF excimer laser beam. There is always a certain degree of tapering during the ablation, which can be measured experimentally and corrected for.

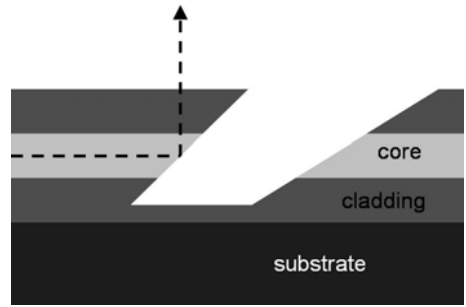

**Figure 2: schematic showing the principle of the TIR mirror.**

The ablated trench contains two interfaces which can both be used to deflect the light beam over 90°. Because of the tapering, only one of the two interfaces will have a 45° angle. In case the core-air interface has a 45° angle, the beam deflection is based on TIR at the polymer-air interface. In the case of the air-core interface two additional processing steps are required. In order to function as a mirror, the facet has to be metal coated and the trench has to be filled with cladding material. We present the results on the TIR mirrors. The schematic is shown in fig. 2.

The surface roughness of this facet can not be measured because of the fact that it is enclosed. The average RMS surface roughness of the other interface is 53nm on a scan area of 52µm×174µm measured with an optical profiler (Wyko NT3000).

Measurements have been carried out on a demo board which contains an array of multimode optical waveguides, integrated on the PCB, and an ablated TIR mirror at a wavelength of 850nm. The light is coupled into the optical waveguides with a multimode optical fiber with a core diameter of 50µm and a numerical aperture (NA) of 0.2. The light that is coupled out vertically at the output facet is detected with a multimode optical fiber with core diameter 100µm and NA 0.29. The Truemode waveguides have NA 0.3. The TIR is not coated with a thin Au layer; this could however increase the coupling efficiency. The first measurement results give an average coupling loss of 3.6dB for the entire link (coupling into the Truemode waveguides, propagation through the waveguide, 90° deflection at the 45° facet and outcoupling at the output facet towards the fiber). The vertical distance between the output facet and the fiber has however not been optimized. Optimization of the ablation parameters (pulse energy and repetition rate) and the use of a Au coating could however improve the efficiency and will be investigated in the near future.

**Resonant grating coupler**

The solution proposed by Laboratoire Hubert Curien is the use of a resonant grating structure at the bottom of the guiding layer as coupling element, in the place of a 45° micro-mirror. Such a structure can theoretically diffract more than 90% of the incident light in the 1st diffraction order [3], which angle can be adjusted to couple light into the optical waveguide. The resonance effect is obtained by combining a metallic diffraction grating and a thin dielectric layer [4] which refractive index is higher than the one of the covering multimode waveguide. The whole component is made thanks to standard planar technologies.

The component includes two resonant gratings (Fig. 3), 3 cm apart, to couple light respectively towards and from the optical waveguide. The grating corrugation in the pyrex substrate is obtained by first exposing a 300 nm thick positive resist to an interferogram at the HeCd laser wavelength of 442 nm, then by reactive ion beam etching into the pyrex surface. The two gratings are made simultaneously and are rigorously parallel. Then a 150 nm thick gold layer is evaporated on the corrugated parts of the substrate in order to create the metallic reflecting grating and a 176 nm thick $HfO_2$ layer is deposited by sputtering on the gold layer to complete the resonant structure. Finally the whole pyrex plate is dip-coated with a 50 µm PMMA layer which forms the multimode waveguide.

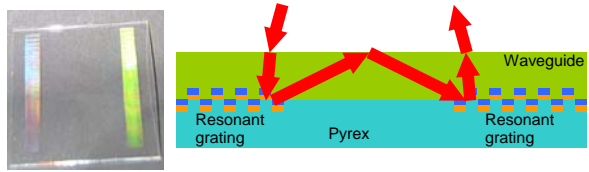

**Figure 3: Left: top view of the 5x5mm² pyrex plate on which two resonant gratings are etched (colored bands) to couple the light to and out the waveguide deposited on the whole plate. Right: sketch of the cross-section of the component with two coupling gratings.**

The angular width of the designed resonance is few degrees large and allows maintenance of high diffraction efficiency with a focused beam. The measurement set-up used for the optical characterization of the component is sketched on Fig. 4. The laser wavelength is 850 nm, the coupling grating is placed in the focal plane of a lens of 60 mm focal length and the main detector measures the intensity of the light diffracted by the out coupling grating. A reference detector records simultaneously the fluctuations of the incident light. A rotation stage not drawn on the sketch allows us to vary the incidence angle of the focused beam on the coupling grating. In such a configuration the overall efficiency of the component, ie the ratio of the out coupled power and the incident power, reaches more than 60% for incidence angles in the range [-3°; -3.5°]. This means that the overall losses, which include the light reflected at the air-waveguide interfaces, the light absorbed by the metal layer, the light scattered during the propagation in the multimode waveguide and on the diffracting structures, do not exceed 2.2 dB.

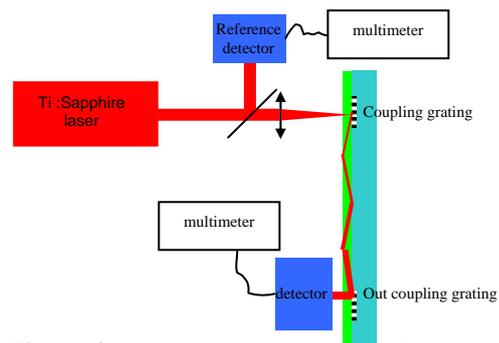

**Figure 4: measurement set-up used to characterize the component.**

Some remarks have to be add to show the versatility of such a grating coupler. It could also be used to couple free space light towards channel waveguides perpendicular to the grating lines. In order to simplify the making process of the grating, the latter can be made without etching by using a photosensitive hybrid sol-gel thin film deposited on the pyrex substrate [5].

**Pluggable out-of-plane coupler**

At the Vrije Universiteit Brussel, we investigate the use of a pluggable micro-optical component to couple the light to or from PCB-integrated waveguides by inserting it into a micro-cavity fabricated in the board. The insert contains a 45° micro-mirror which is used to deflect the light beam over 90° and which can be Au-coated to increase the coupling efficiency. For the fabrication of the pluggable out-of-plane coupler, we make use of our in-house rapid prototyping technology of Deep Proton Writing (DPW) [6]. It consists of the following processing steps, as illustrated in Fig. 5. First a collimated 8.3MeV proton beam is used to irradiate an optical grade PMMA sample according to a predefined pattern by translating the PMMA sample, changing the physical and chemical properties of the material in the irradiated zones. As a next step, a selective etching solvent is applied for the development of the irradiated regions. This allows for the fabrication of (2D) arrays of micro-holes, optically flat micro-mirrors and micro-prisms, as well as alignment features and mechanical support structures. On the other hand, an organic monomer vapour can be used to expand the volume of the bombarded zones through an in-diffusion process. This enables the fabrication of spherical (or cylindrical) micro-lenses with well defined heights. If necessary, both processes can be applied to different regions of the same sample, yielding micro-optical structures combined with monolithically integrated micro-lenses.

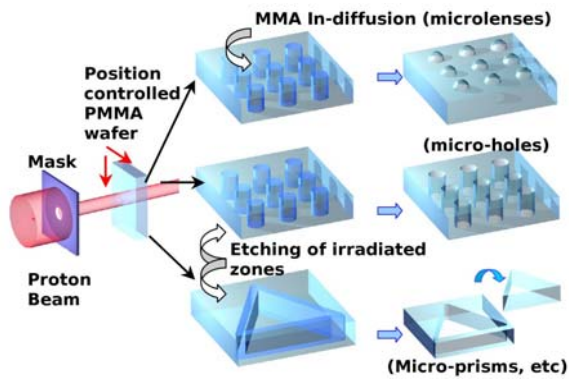

**Figure 5: Deep Proton Writing: basic processing steps. Proton beam exposure of PMMA is followed by selective etching and/or swelling**

We use high molecular weight PMMA with a thickness of 500μm, which allows the 8.3MeV protons to completely traverse the sample. During the irradiation step, the PMMA sample is semi-continuously translated perpendicularly to the beam in steps of 500nm using high-precision translation stages with an accuracy of 50nm, according to the pre-defined pattern shown in the bottom part of Fig. 6. Optimal surface roughness results are obtained by using a proton dose of 50pC per step of 500nm, with a proton current of 160pA. This current is monitored by measuring the charge that the protons induce on a target located directly behind the sample. The deposited dose can then be determined by integrating this proton current during the irradiation using a precision-switched integrator trans-impedance amplifier that aims at compensating any current fluctuations of the proton source.

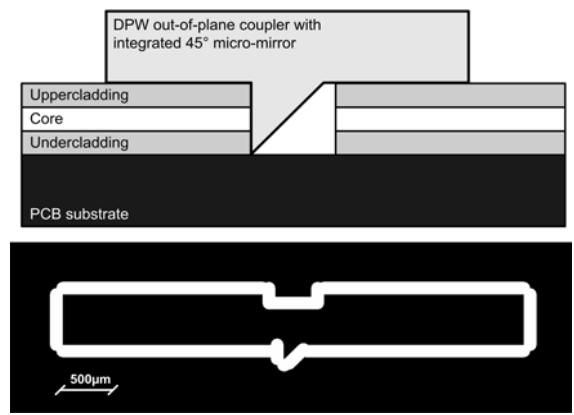

**Figure 6: Schematic operation principle (top) and design lay-out (bottom) of a pluggable out-of-plane coupling component**

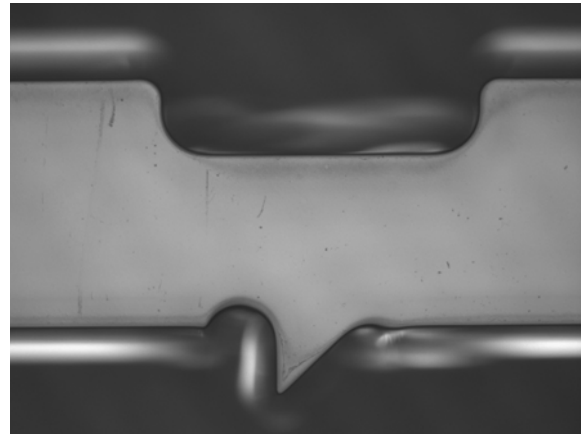

**Figure 7: Fabricated out-of-plane coupling component with integrated 45° micro-mirror**

After the exposure step, the irradiated zones can be selectively etched, resulting in micro-components with high-quality optical surfaces. An optical microscope picture of a fabricated pluggable out-of-plane coupler is given in Fig. 7.

It is obvious however that DPW is not a mass fabrication technique as such. However, one of its assets is that, once the master component has been prototyped with DPW, a metal mould can be generated from the master by applying electro-plating. After removal of the plastic master, this metal mould can be used as a shim in a final micro-injection moulding or hot embossing step [7]. This way, the component can be mass-produced at low cost in a wide variety of high-tech plastics.

For the characterization of the critical optical surfaces of the component, namely the flat top exit facet and the 45° mirror facet, we use a WYKO NT-2000 non-contact optical surface profiler (Veeco). Since the entrance facet is not accessible with the microscope objective, this surface was not measured, but its surface roughness will be analogous to the two others. The surface roughness analysis reveals that the flat top part has an average local RMS surface roughness $R_q$ of 14.1nm ± 2.7nm measured over an area of 60μm by 46μm. We averaged at least 5 measurements of randomly chosen positions. Applying the same measurement method to the 45° angled facet reveals an RMS roughness of 17.1nm ± 4.2nm. The flatness $R_t$ or peak-to-valley difference along the depth of 500μm of the component is measured to be smaller than 2.5μm. This is due to the scattering of the protons during the interaction with the PMMA. As a conclusion, we can say that our developed DPW surfaces have a very good and reproducible optical quality: almost flat and a very low RMS roughness.

Measurements have been carried out on an optical board, containing Truemode™ optical waveguides and a laser ablated micro-cavity. The out-of-plane coupler is inserted passively into the micro-cavity. Again, 850nm light is coupled into the waveguides using a MMF with core diameter 50μm and NA 0.2 and the power coupled out by the DPW pluggable coupler is detected by means of a MMF with core diameter 100µm and NA 0.29. The coupling efficiency that is measured for the entire link is -5.68dB. It should be noted that when the DPW out-of-plane coupler is measured in a direct fiber-to-fiber coupling scheme, using the same input and output MMF as described above, efficiencies up to -0.72dB are measured.

Another important advantage of the pluggable out-of-plane coupler fabricated with DPW is the fact that micro-lenses can be monolithically integrated in the component, and that the concept can easily be extended towards multilayer optical waveguide structures integrated on printed circuit boards, as reported in [8] and [9].

**Glass micro-mirror**

At VTT, micro-mirrors for 90° beam turning were fabricated by grinding and polishing one edge of 100µm thick glass substrate in such as way that a 45° bevel was formed. Several substrates were grinded by stacking them into a jig which made it possible to mount them in a 45° angle precision lapping and polishing machine. After polishing, the substrates were diced with a dicing saw. An aluminum coating layer was evaporated on the glass surfaces for high reflectivity. Also, a thin layer of chrome was evaporated in order to achieve better adhesion of the aluminum coating.

Optical loss measurements have been carried out on an optical board with an embedded glass mirror. The top view of the demoboard is shown in Fig. 8. The measurement set-up is schematically shown in Fig. 9. MMF with core diameter 50µm and NA 0.22 is used to couple light into the multimode optical waveguides. The light beam that is deflected by the glass mirror is detected with a MMF with core diameter 200µm and NA 0.22, which is equipped with a coupling lens. Excess loss caused by the micro-mirror coupling was estimated from the measured losses by subtracting the waveguide loss and the fiber-coupling losses, which were obtained from the measurements done using butt-coupling to waveguides of the same length at both ends. The excess loss calculated thus includes losses due to the quality (roughness, angle) of the mirror, due to refraction and scattering from the under cladding surface, and due to the possibly poorer quality of the lithographically patterned waveguide facet compared to a sawed facet.

Four channels (i.e. waveguides) were measured on two different demo boards, resulting in eight measurement results. The excess losses were between 4.0dB and 6.5dB with an average of 5.1dB [10]. The variation is partly due to the varying quality of the sawed, not polished, waveguide facet at the in-coupling. It should be also pointed out that the excess loss is probably slightly different in the fully assembled demonstrator, since the micro-lens based collecting optics performs differently than the fiber-coupling optics used in this experiment. It is clear that the edge of the under-cladding layer blocks part of the optical beam between the waveguide end facet and the micro-mirror. This part of the optical power is lost due to scattering and reflections to wrong directions, thus significantly increasing the total path loss.

The average RMS surface roughness of the mirror facet, measured on a scan area of 59µm×45µm, is 42.3nm. The coupling efficiency may be improved by placing the mirror into a micro-cavity in the optical layer. The distance between the output facet of the waveguides and the mirror facet can in this way be decreased. Tests where the mirror is placed in the same demo board as the pluggable out-of-plane coupler will be carried out in the near future.

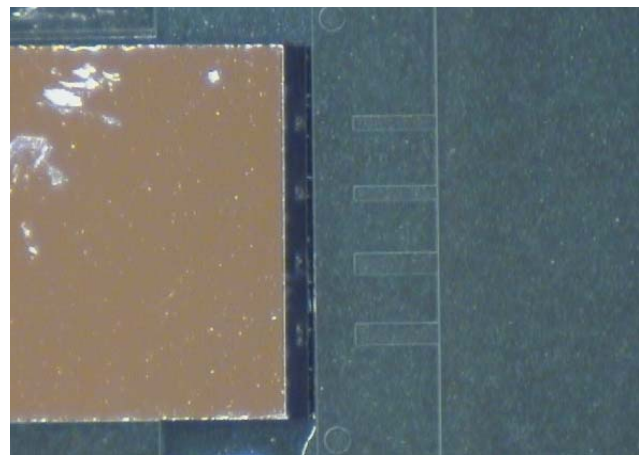

**Figure 8: top view of the demo board with the inserted glass mirror.**

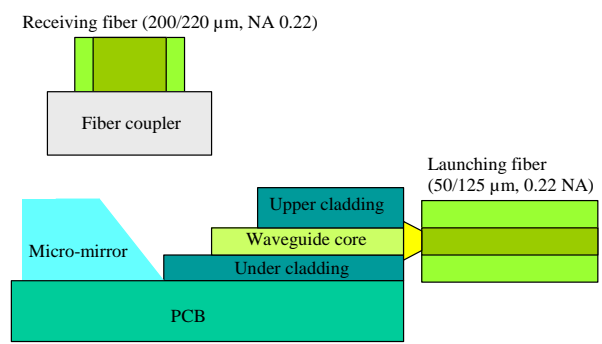

**Figure 9: measurement set-up used to characterize the excess loss.**

**Conclusion**

Main advantage of the ablated micro-mirrors is the fact that they are integrated with the optical waveguides. We can in this way achieve high alignment accuracies with use of passive alignment. The performance of these mirrors can be improved by optimizing the ablation parameters and with the application of a Au coating on the mirror facet. This way, the trench could be filled with cladding material to protect the mirror from dust and moist intrusion.

The advantage of the grating coupler is the very high coupling efficiency that can theoretically be achieved. Drawback is the fact that the device is wavelength dependent, and the strong requirements on the input side (input angle, low beam divergence).

The pluggable out-of-plane coupler is very versatile because it can easily be extended towards multilayer structures, where multiple optical layers are stacked. It is also possible to monolithically integrate micro-lenses which can ease the alignment tolerances. Although PMMA is not suitable for standard PCB processing, the DPW prototypes are compatible with low-cost mass-fabrication techniques in high-tech plastics capable of withstanding high temperatures.

The glass mirrors have excellent optical and thermal properties and are fully compatible with standard PCB manufacturing and soldering processes. We believe that the efficiency can be improved by putting the output facet of the waveguides closer to the mirror facet.

**Acknowledgements**

Nina Hendrickx would like to acknowledge the Institute for the Promotion of Innovation by Science and Technology in Flanders (IWT Flanders) for financial support. Jürgen Van Erps acknowledges the FWO for financial support. This work was carried out within the framework of the network of excellence on micro-optics (NEMO), supported by the European Commission through FP6 program.